\documentclass[superscriptaddress,prl, amsmath,amssymb,reprint,showkeys,showpacs]{revtex4-1} %%%Dos Columnas

\usepackage{graphicx}
\usepackage{graphics}
\usepackage{amssymb}
\usepackage{amsmath,bm}
\usepackage{float}
\usepackage[usenames,dvipsnames]{color}
\usepackage[export]{adjustbox}
\usepackage{hyperref}
\usepackage{gensymb}

\usepackage{soul,xcolor}
\setstcolor{red}

\hypersetup{
    colorlinks=true,
    linkcolor=blue,
    citecolor=blue,
    filecolor=cyan,      
    urlcolor=cyan,
}
\newcommand{\ket}[1]{\left|  #1 \right\rangle}

\newcommand{\aver}[1]{\ensuremath{\langle {#1} \rangle}}

\begin{document}

\title{Strongly Correlated Quantum Gas Prepared by Direct Laser Cooling}
%\title{Fast Preparation of a Strongly Correlated Quantum Gas by Direct Laser Cooling}
%\title{Cooling Cesium atoms to quantum degeneracy with resonant light}

\author{Pablo Solano}
\affiliation{Department of Physics, MIT-Harvard Center for Ultracold Atoms, and Research Laboratory of Electronics, Massachusetts Institute of Technology, Cambridge, Massachusetts 02139, USA}

\author{Yiheng Duan}
\affiliation{Department of Physics, MIT-Harvard Center for Ultracold Atoms, and Research Laboratory of Electronics, Massachusetts Institute of Technology, Cambridge, Massachusetts 02139, USA}

\author{Yu-Ting Chen}
\affiliation{Department of Physics, MIT-Harvard Center for Ultracold Atoms, and Research Laboratory of Electronics, Massachusetts Institute of Technology, Cambridge, Massachusetts 02139, USA}
\affiliation{Department of Physics, Harvard University, Cambridge, Massachusetts, 02138, USA}

\author{Alyssa Rudelis}
\affiliation{Department of Physics, MIT-Harvard Center for Ultracold Atoms, and Research Laboratory of Electronics, Massachusetts Institute of Technology, Cambridge, Massachusetts 02139, USA}

\author{Cheng Chin}
\affiliation{James Franck Institute, Enrico Fermi Institute, Department of Physics, University of Chicago, Chicago, Illinois 60637, USA}

\author{Vladan Vuleti\'{c}}
 \email{Corresponding author email: vuletic@mit.edu}
\affiliation{Department of Physics, MIT-Harvard Center for Ultracold Atoms, and Research Laboratory of Electronics, Massachusetts Institute of Technology, Cambridge, Massachusetts 02139, USA}

\date{\today}

\begin{abstract}

We create a one-dimensional strongly correlated quantum gas of $^{133}$Cs atoms with attractive interactions by direct laser cooling in 300~ms. After compressing and cooling the optically trapped atoms to the vibrational ground state along two tightly confined directions, the emergence of a non-Gaussian time-of-flight distribution along the third, weakly confined direction reveals that the system enters a quantum degenerate regime. We observe a strong reduction of two- and three-body spatial correlations and infer that the atoms are directly cooled into a highly correlated excited metastable state, known as a super-Tonks-Girardeau gas.

\end{abstract}
\maketitle

%%%%%%%%%%%%%%%%%%%%%%%%%%%%%%%%%%%%%%%%%%%%%%%%%%%%%
Laser trapping and cooling techniques enable the preparation of atomic ensembles at ultracold temperatures, where quantum effects dominate \cite{Phillips1998,Dalibard1989}. However, exclusively using optical cooling to reach the quantum degenerate regime is challenging: re-scattering of cooling light inside the optically dense atomic ensemble causes excess recoil heating \cite{Castin1998}, while atomic collisions in the presence of light can lead to heating or molecule formation \cite{Burnett1996}. For these reasons, the standard final step to quantum degeneracy is evaporative cooling \cite{Davis1995, Anderson1995,Bradley1995,DeMarco1999,Lu2011}. However, evaporation is relatively slow, relies on favorable atomic collisional properties, and requires substantial atom loss.

Schemes based on laser cooling alone have recently succeeded in reaching quantum degeneracy \cite{Stellmer2013,Hu2017a,Urvoy2019}. Common to those techniques is the reduction of detrimental effects of the cooling light, either by shielding the densest region from the light (for $^{84}$Sr) \cite{Stellmer2013} or by using far-off-resonance cooling light (for $^{87}$Rb) \cite{Hu2017a,Urvoy2019} to reduce light-induced collisions. In both cases, the system was in a weakly correlated state with repulsive two-body interactions (scattering length $a>0$), such that the condensate is stable against collapse.

Among the alkali atoms, $^{133}$Cs is notoriously difficult to evaporatively cool. It features large two- and three-body inelastic collision rates \cite{Leo2000,Chin2000,Kraemer2006}, while its large negative scattering length \cite{Vuletic1999,Leo2000,Chin2000,Chin2010} and associated strong attractive atom-atom interaction result in collapse of a three-dimensional condensate. Condensation of $^{133}$Cs was eventually achieved through slow evaporation at low atomic density in a combination of optical and magnetic traps \cite{Weber2003} using a magnetic Feshbach resonance to tune the scattering length to a positive value \cite{Vuletic1999,Leo2000,Chin2010}. 

\begin{figure}
\includegraphics[width=0.48\textwidth]{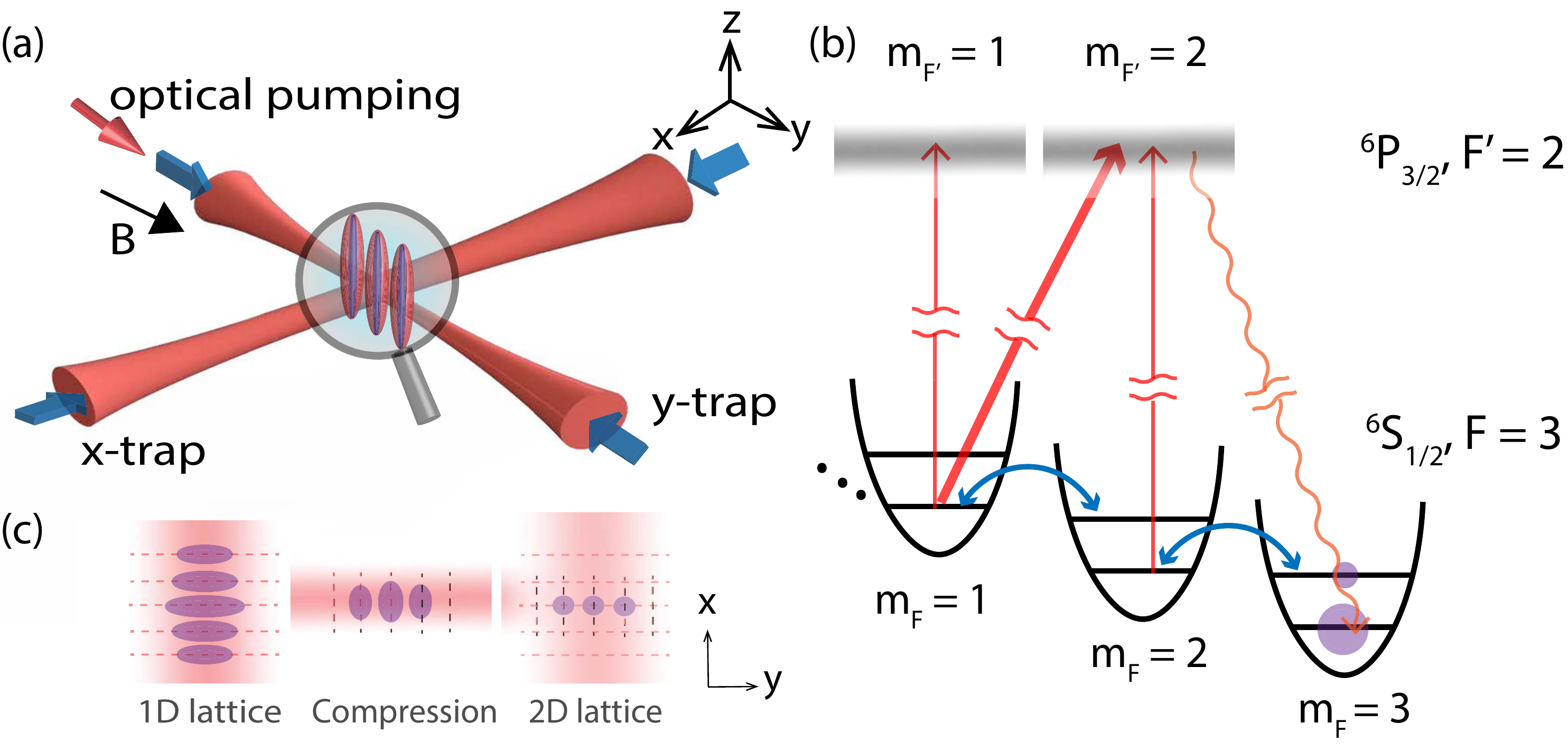}
\caption{Experimental setup. (a) Two crossed optical standing waves create a lattice of highly anisotropic cigar-shaped traps. An optical-pumping beam is applied along the $y$-axis, with a magnetic field at a small angle ($\alpha \simeq 10^{\circ})$ to the $y$-axis. (b) Atomic level structure and cooling procedure. The red arrows represent optical pumping, and the blue arrows represent Raman transitions driven by the trap light. The Raman transition removes one quantum of vibrational energy, while the optical pumping restores the initial internal state. (c) By varying the powers of the two trapping beams in combination with laser cooling, the atoms are compressed into a small number of traps, where a final cooling yields a sTG gas.}
\label{setup}
\end{figure}

A particularly interesting situation arises when the degenerate gas is in a strongly correlated state that cannot be described by a mean-field theory.
Such a regime is reached, e.g., when a quantum gas with strong repulsive interactions is so tightly confined in two directions that it becomes effectively one-dimensional (1D) \cite{Moritz2003,Stoferle2004,Tolra2004, Kinoshita2004,Paredes2004}. Then at sufficiently low linear density, the atoms avoid each other, effectively behaving like fermions in a regime known as a Tonks-Girardeau gas \cite{Olshanii1998, Dunjko2001,Tolra2004, Kinoshita2004,Paredes2004,Moritz2003}. Surprisingly, this behavior persists even at large negative scattering length, in spite of the strong attraction, where the atoms enter a strongly correlated metastable many-body state \cite{Astrakharchik2004,Astrakharchik2005,Kormos2011,Haller2009}. Such a `super-Tonks-Girardeau' (sTG) gas was previously generated in a pioneering experiment by transferring a Bose-Einstein condensate adiabatically into a two-dimensional (2D) optical lattice \cite{Haller2009}.

In this Letter, we demonstrate direct laser cooling of a quantum gas with attractive interactions into a strongly correlated sTG state that cannot be described by mean-field theory. The motion along two tightly confined directions $x,y$ is continuously cooled to the quantum ground state by degenerate Raman sideband cooling (dRSC) \cite{Vuletic1998,Hamann1998,Kerman2000,Han2000,Bouchoule2002}. Along the third, weakly confined $z$ direction, atomic collisions transfer energy to the tightly trapped directions. After compression of the gas of $^{133}$Cs atoms into a small array of optical traps and cooling, a non-Gaussian momentum distribution along $z$ emerges, evidencing the onset of quantum degeneracy in this quasi-1D system with large negative scattering length $a = -130$~nm  \cite{Leo2000,Chin2000}. Measurements of inelastic collisions show that spatial two- and three-body correlations are suppressed by almost one and two orders of magnitude, respectively. We infer that the system is directly cooled into a metastable excited gas-like state (sTG gas), stabilized by the strong attractive interaction \cite{Astrakharchik2004}. Using resonant light for the optical pumping process in dRSC, the sTG gas is prepared in less than 300 ms, more than ten times faster than previous condensation of $^{133}$Cs by evaporative cooling~\cite{Weber2003,Hung2008}.

The process begins by loading $^{133}$Cs atoms from a magneto-optical trap into a standing-wave trap operating at a wavelength $\lambda=1064$~nm ($x$-trap, beam waist $w_x=17 \mu$m). We then perform dRSC on the 2D gas, optically pumping the atoms into the lowest-energy hyperfine and magnetic sublevel $\ket{F=3,m=3}$ of the electronic ground state $^6S_{1/2}$. Trapping light with finely tuned polarization drives Raman transitions transferring atoms to the neighboring magnetic sublevel $\ket{F=3,m=2}$ while reducing the vibration quantum number by one. Optical pumping via the $^6P_{3/2}, F'=2$ hyperfine manifold back into $\ket{F=3,m=3}$ continuously removes entropy from the system. (See Fig. \ref{setup}b for the atomic level structure and Supplemental Materials (SM) \cite{SM} for details on trap parameters and cooling procedure.)

\begin{figure}
\includegraphics[width=0.4\textwidth]{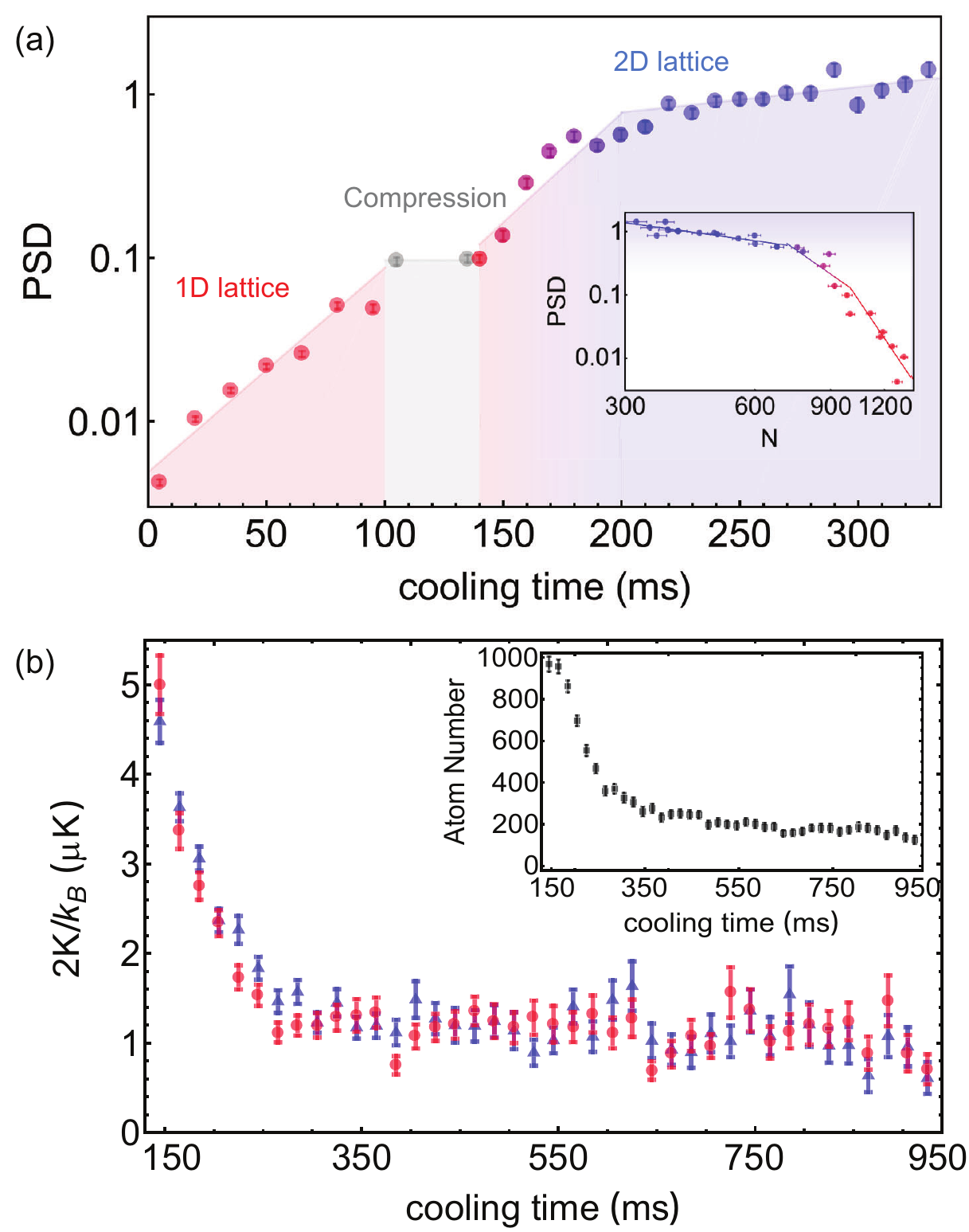}
\caption{(a) Phase space density during the cooling sequence. Following a pre-cooling stage (red region), the atoms are compressed in 40 ms into fewer traps (see text). During the final cooling stage the system crosses over into a quantum degenerate region. The inset shows the efficiency of the cooling, displaying the PSD against the remaining atom number. The blue shaded region in the inset represents the PSD at which quantum degeneracy becomes observable. (b) Kinetic energy $K$ of the atoms after sudden release from the trap in a time-of-flight measurement, as a function of cooling time, for the weakly confined (red dots) and strongly confined (blue triangles) directions. The inset shows the atom number as a function of cooling time.}
\label{cooling}
\end{figure}

After 100 ms of cooling, the 2D gas reaches a temperature of $T=2.5 \mu$K and peak classical phase space density (PSD) of PSD$\sim0.1$ (see SM \cite{SM} for the exact definition). If the cooling were to continue in this geometry, the PSD would subsequently decrease due to light-induced atom loss \cite{Burnett1996}. We then turn on a second lattice trap ($y$-trap, waist $w_y=6.5 \mu$m) transverse to the first one. (see Fig. \ref{setup}a). This configuration creates a 2D array of elongated cigar-shaped traps along the $z$ direction. Immediately after switching on the $y$-trap we adiabatically turn the $x$-trap off and back on to remove atoms not confined to the overlap region of the two traps (see SM for details \cite{SM}). This prepares $N \simeq 1000$ atoms distributed in an array of cigar-shaped traps with root-mean-square size of $1.3 \times 2.5$ traps, and a peak occupation of $N_1 \simeq 50$ atoms per trap, as depicted in Fig. \ref{setup}c. Due to the spatial compression when turning off the $x$ trap, the atoms' temperature is increased to $T \simeq 5 \mu$K, at constant PSD (see Fig. \ref{cooling}a).

The final cooling stage employs two-dimensional dRSC along $x$ and $y$. The trapping frequencies are $\omega_{x,y}=2 \pi \times 50$ kHz in the transverse directions and $\omega_{z}=2 \pi \times 2.9$ kHz along the weakly confined direction. After 120 ms of cooling, the 2D ground state in the $xy$-plane is reached (see Fig. \ref{cooling}b). During cooling, atoms are lost at a moderate rate due to light-assisted inelastic collisions; once the atoms are cooled to the 2D ground state, the loss rate is substantially reduced, presumably due to the lower cooling and associated photon scattering rate. The two-body loss acts to equalize the populations in different traps, and after 200~ms of two-dimensional dRSC we have 50 mostly equally filled traps with $N_1=6$ atoms in each, for a total atom number $N \simeq 300$ (see SM). The local peak density is $n_0=1.1 \times 10^{15}$cm$^{-3}$.

The evolution of the classical PSD throughout the cooling sequence is shown in Fig. \ref{cooling}a. The cooling efficiency in the presence of atom loss can be characterized by the logarithmic slope of PSD increase to atom number loss,  $\eta=-\frac{d(\text{ln}\text{PSD})}{d(\text{ln}N)}$. During the pre-cooling stage, up to PSD$\sim$0.5, we observe a very high efficiency $\eta=10\pm0.3$ (see inset of Fig. \ref{cooling}a), whereas a typical evaporative cooling has $\eta \sim 3-4$, with the highest reported values $\eta\simeq 6$ \cite{Olson2013,Hung2008}. For values PSD$\gtrsim 0.5$, the classical PSD no longer coincides with the (higher) occupation per quantum state, and should only be regarded as a qualitative measure of cooling efficiency.

\begin{figure}
\includegraphics[width=0.48\textwidth]{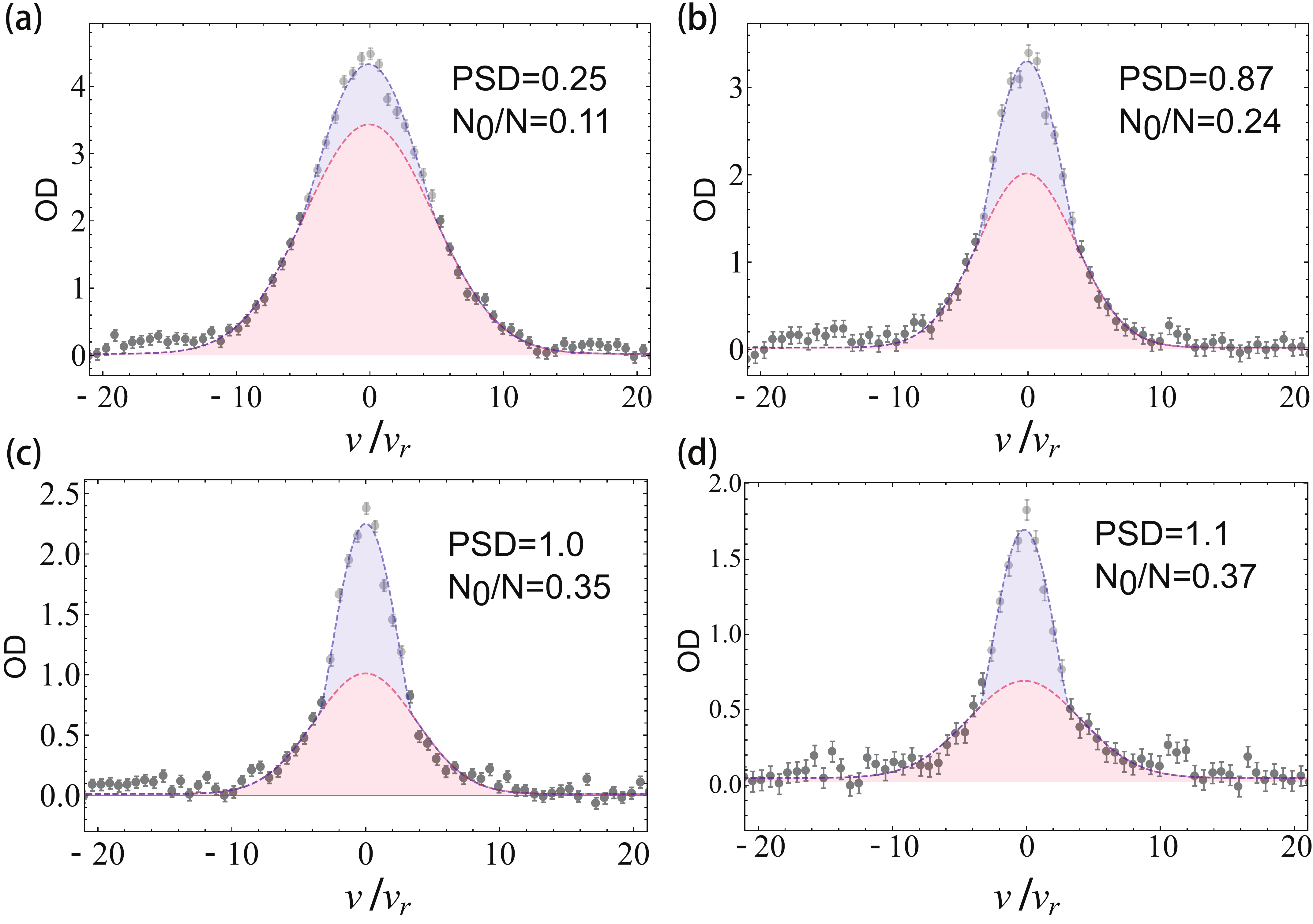}
\caption{Velocity distribution of the atoms after sudden release from the trap, normalized by the recoil velocity. Each plot is an average of 800 TOF images. (a)-(d) correspond the momentum distribution along the direction of weak confinement after 20, 100, 200, and 400 ms of cooling respectively. The red area represents the fit of a Gaussian distribution to the wings, which we defined to be one standard deviation away from the peak, and the blue area represents the fit of the data minus the Gaussian distribution to a parabola. The value of $N_0/N$ is determined from the ratio of the blue over the total area. For (d), the reduced $\chi^2$ of the fit to a single Gaussian is $\chi^2>12$, while for the Gaussian distribution plus an inverted parabola it is $\chi^2=1.17$.}
\label{fits}
\end{figure}

During the final cooling stage (Fig. \ref{cooling}b), a non-Gaussian momentum distribution emerges along the $z$ direction in a time-of-flight (TOF) measurement (Fig. \ref{fits}), indicating the crossover to quantum degeneracy in this effectively 1D system. The distribution can be well captured by a bimodal fit. We characterize the degree of quantum degeneracy of the system by the ratio $N_0/N$ of the area $N_0$ under a central peak that deviates from a Gaussian distribution and the total area under the TOF curve $N$ (see SM for details \cite{SM}). We observe a maximum ratio $N_0/N \simeq 37$\% after 400 ms of cooling (see Fig. \ref{fits}d). The momentum distribution in the tightly confined direction, on the other hand, always shows a Gaussian distribution, as expected for zero-point motion with $k_B T < \hbar \omega_{x,y}$ (see SM \cite{SM}). 1D systems exhibit a smooth crossover to quantum-degeneracy \cite{Petrov2000,Bouchoule2007}, and we observe a small quantum degenerate component even for PSD$<$1.

The cooling prepares the system in a strongly correlated, effectively 1D state that survives at large negative scattering length and high density, and is largely immune to two-body radiative losses and three-body recombination loss. Effectively 1D systems can be characterized by the combination of linear density $n_{\text{1D}}$ and 1D scattering length, given by $a_{\text{1D}}=-a_{\perp}\left(\frac{a_{\perp}}{a}-C\right)$, where $a_{\perp}=\sqrt{\hbar
/m \omega_{x,y}}$ is the ground-state size in the transverse directions, $m$ the atomic mass, $a$ the three-dimensional scattering length, and $C \approx 1.0326$ a constant \cite{Olshanii1998}. The strength of the interactions is characterized by the dimensionless parameter $\gamma=2/(n_{\text{1D}} |a_{\text{1D}}|)$ that can be interpreted as the ratio of interaction energy to kinetic energy \cite{Paredes2004,Kinoshita2004}. $\gamma \ll 1$ corresponds to a weakly interacting Bose gas, while $\gamma \gg 1$ describes a strongly interacting Tonks-Girardeau gas (sTG gas) for $a_{\text{1D}}<0$ ($a_{\text{1D}}>0$).
%Note that a lower density $n_{\text{1D}}$ corresponds to a more strongly interacting system.
For our parameters with three-dimensional scattering length $a = -130$~nm and $|a| > a_{\perp}= 39$~nm, the system is in the unitary regime with $a_{\perp} \approx a_{\text{1D}} = 52$~nm. Assuming a Gaussian density distribution along the tubes, these parameters correspond to a value $\gamma_0=8$ at the trap center in the local density approximation, with an estimated systematic uncertainty of a factor of 2. A sTG gas has been predicted to be stable against attraction-induced collapse for $\gamma>5.7$ \cite{Astrakharchik2004}. The near-coincidence of this value with our observed $\gamma_0$ may indicate that the number of atoms in our traps is limited by the stability condition of the sTG gas.

Similar to the Tonks-Girardeau gas, the sTG gas is a highly correlated state where the spatial wavefunction of the bosons is `fermionized.'
The associated suppression of two- and three-body short-range correlations $g^{(2)}$ and $g^{(3)}$ has been previously observed for the Tonks gas with repulsive interactions \cite{Haller2011}, but not for the sTG gas. In the following, we investigate experimentally this suppression $g^{(2)}$, $g^{(3)} < 1$ that is expected to persist even at finite temperature \cite{Astrakharchik2004,Kormos2011}.

\begin{figure}
\includegraphics[width=0.4\textwidth]{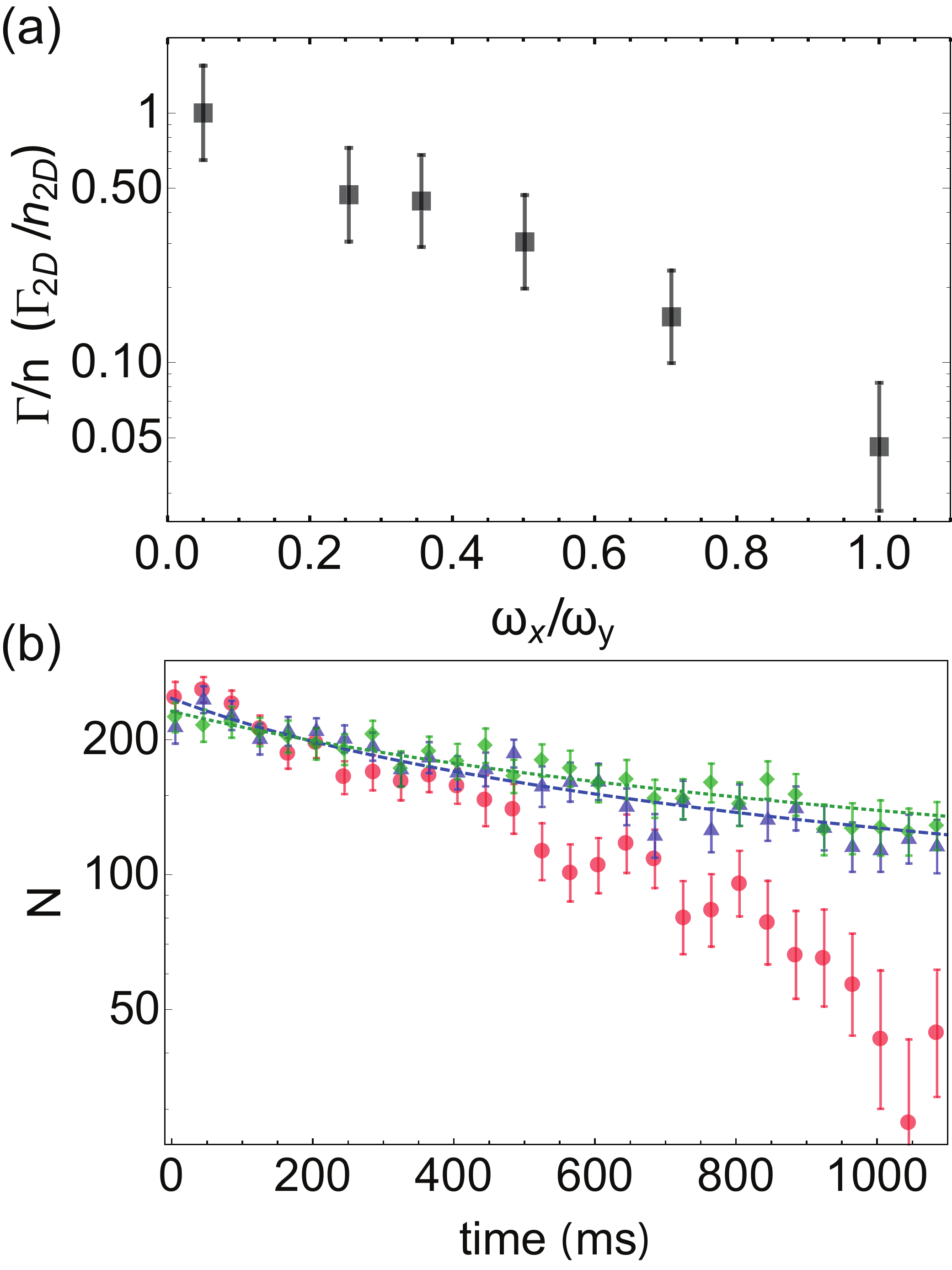}
\caption{(a) Density-normalized two-body loss rate as a function of the dimensionality of the gas in the presence of cooling light. $\omega_x$ is being varied at fixed $\omega_y/(2\pi)=50$~kHz between $\omega_x/\omega_y=0$ and $\omega_x/\omega_y=1$.
The temperature is constant at 1.2 $\mu$K. (b) Three-body loss in the absence of cooling light for a 2D gas (green diamonds, $\omega_x/\omega_y=0$), a 1D gas (blue triangles, $\omega_x/\omega_y=1$), and in between (red circles, $\omega_x/\omega_y=0.25$. The dotted lines are fits to three-body loss. The average temperature is $T=3.5 \mu$K. For the 1D gas, the density is 13 times higher than for the 2D gas, while the three-body loss rate is only increased by a factor of 1.4. The atom number evolution in the intermediate regime between 1D and 2D, $\omega_x/\omega_y=0.25$, cannot be described by three-body loss.}
\label{lossrate}
\end{figure}

The two-body short-range atom-atom correlation function $g^{(2)}$ can be probed with light-assisted loss \cite{Kinoshita2005b}. The loss rate constant $\Gamma$ with $\dot{N} = -\Gamma N$, is given by $\Gamma = G g^{(2)} \aver{n}$, where $G$ is the light-induced loss rate constant set by the light intensity and detuning, and $\aver{n}$ is the average atomic density. While $G$ is difficult to evaluate from first principles, we can keep the laser power and detuning constant while changing the dimensionality of our system. To this end, we reduce the $x$ confinement by a variable factor after quantum degeneracy in the cigar-shaped traps has been reached. Fig. \ref{lossrate}a shows the observed ratio $\Gamma/\aver{n} =G g^{(2)}$ as a function of vibration frequency ratio $\omega_x/\omega_y$. The loss rate constant $\Gamma$ is measured, and $\aver{n}$ is calculated from the measured temperature, atom number $N$, and trap vibration frequencies. Compared to a 2D gas ($\omega_x/\omega_y=0$), the density-normalized light-induced loss rate $\Gamma/n$ is substantially reduced for the 1D gas ($\omega_x/\omega_y=1$) by a factor 20. Part of that change can be attributed to the change of $G$ with dimensionality: Pairs of atoms are excited by the light near the Condon point $r_C$ in interatomic distance \cite{Burnett1996}, which for resonant light of wavelength $\lambda$ is at $r_C \sim \lambda/(2\pi)$, while the atoms are confined to a smaller distance scale $a_\perp$ in the direction of tight confinement. Consequently we expect a reduced loss rate constant $G_{1D}$ compared to $G_{2D}$ by a factor $G_{1D}/G_{2D} \approx a_\perp/r_C =0.3$ (see SM). Taking this into account, we find from our measurements $g^{(2)}_{2D}/g^{(2)}_{1D} \approx 5$, i.e. $g^{(2)}_{1D} \approx 0.4$, in agreement with the theoretically expected value $g^{(2)}_{1D} = 0.4$ for a sTG gas with $\gamma = 8$ at our temperature \cite{Kormos2011}. 

Three-body correlations $g^{(3)}$ can be measured by turning off the cooling light after reaching the quantum degenerate regime, and observing the atom number evolution as a function of dimensionality between $\omega_x/\omega_y=0$ (2D gas) and $\omega_x/\omega_y=1$ (1D gas). We measure the atom loss vs. time and fit the data to the functional form for three-body loss (see Fig. \ref{lossrate}, and SM \cite{SM} for details). Comparing the 2D and 1D gases, we observe a suppression of three-body recombination rate constant by a factor of $120 \pm 30$. ($K_{\text{1D}}=(4\pm2)\times10^{-28}\text{~cm}^6\text{s}^{-1}$ and $K_{\text{2D}}=(5\pm3)\times10^{-26}\text{~cm}^6\text{s}^{-1}$ for the 1D gas and 2D gas, respectively. The 2D case is in agreement with theoretical predictions for a 3D gas~\cite{Eismann2016}). This provides strong evidence that the 1D nature of the system is protecting the dense gas from three-body loss. The theoretically expected value for our interaction parameter $\gamma$ and temperature is $g^{(3)}_{2D}/g^{(3)}_{1D} \approx 100$~\cite{Kormos2011}, close the measured value.

Both for the 1D and the 2D gas we observe a time dependence that is well fitted by three-body decay (Fig. \ref{lossrate}b). Surprisingly, for an intermediate regime between 1D and 2D  (red circles), the atom loss does not follow the characteristic behavior of three-body loss, but the loss speeds up at late times, and very few atoms survive. We hypothesize that the rapid atom loss is due to collapse of the gas at high atomic density in the 1D-2D crossover region when the dimensionality of the system no longer protects the gas against attraction-induced collapse.

In conclusion, we have demonstrated direct laser cooling into a strongly interacting metastable excited phase, a sTG gas. The method is fast and robust, preparing a strongly correlated quantum gas in 300 ms, and enabling measurements with high signal-to-noise ratio. In the future, it will be interesting to tune the scattering length by means of a Feshbach resonance \cite{Vuletic1999,Leo2000,Chin2000,Chin2010}, and expand previous studies of Tonks-Girardeau \cite{Paredes2004,Kinoshita2004} and sTG \cite{Haller2009} gases with improved signal-to-noise ratio. The demonstrated scheme presents a promising tool for reaching quantum degeneracy in atoms with unfavorable collision properties, and can potentially be extended to optically trapped molecules \cite{Anderegg2018}.

\section*{Acknowledgments}

This work was supported by the NSF, the NSF Center for Ultracold Atoms, MURI grants through AFOSR and ARO, and NASA. C.C.  acknowledges support from NSF Grant
No.    PHY-1511696, and the University of Chicago MRSEC, funded by the NSF under Grant
No. DMR-1420709. We are grateful to F. Salces-Carcoba, Alban Urvoy, Zachary Vendeiro, and Martin Zwierlein for stimulating discussions.

\vspace{1cm}

\bibliography{CsCoolingPaperRefs}

\newpage

\begin{widetext}

\section{Supplementary Material: Fast Preparation of a Super-Tonk Gas by Laser Cooling}

\subsection{Experimental Details}
$^{133}$Cs atoms are loaded from a magneto-optical trap (MOT) into a standing-wave trap operated at wavelength $\lambda=1064$ nm with waist $w_x=17\ \mu$m and 100 mW of power ($x$-trap). The lattice is created by a vertically polarized beam and its retroflection with the polarization rotated by about 83\degree, set to obtain trapping frequencies of $\omega_{x}=2 \pi \times 50$ kHz along the lattice and $\omega_{\perp x}=2 \pi \times 1.5$ kHz transverse to the beam propagation, with a calculated trap depth of $U_x/h=2.2$ MHz. After loading into the trap and polarization gradient cooling to a temperature of T=6 $\mu$K, the phase space density (PSD, defined below) is PSD$\simeq4\times10^{-3}$ \cite{Shibata2017}, and the peak density is $n_0\simeq2\times10^{13}\text{cm}^{-3}$. We then perform degenerate Raman sideband cooling (dRSC) of the 2D gas by applying a magnetic field of about 150 mG to match the energy of the $|\nu; 6\text{S}_{1/2},\text{F}=3,\text{m}_{\text{F}}=3\rangle$ and $|\nu-1; 6\text{S}_{1/2},3,2\rangle$ states, where $\nu$ represents the vibrational level of an atom in the direction of tight confinement. In this configuration the trapping light drives the $|\nu;6\text{S}_{1/2},\text{F}=3,\text{m}_{\text{F}}=3\rangle \rightarrow |\nu-1;6\text{S}_{1/2},\text{F}=3,\text{m}_{\text{F}}=2\rangle$ Raman transition, with calculated Rabi frequency of 2$\pi\times$2 kHz. Unlike previous realizations of laser cooling to quantum degeneracy in $^{87}$Rb \cite{Hu2017a,Urvoy2019}, we use %8.5 pW of 
light resonant with the $|6\text{S}_{1/2},\text{F=3}\rangle \rightarrow |6\text{P}_{3/2},\text{F'=2}\rangle $ transition to optically pump the atoms from the $|3,2\rangle$ state back to $|3,3\rangle$ using spontaneous Raman scattering,  removing entropy from the system (see Fig. \ref{setup}b for the atomic level structure). The pump light, with an intensity of 6 $\mu$W/cm$^2$, is mostly $\sigma^+$ polarized with a small component of $\pi$-light, to empty all magnetic sublevels other than the lowest-energy state $|3,3\rangle$. At the end of this first cooling stage after 100 ms, the trap contains $N=4000$ atoms with a peak occupancy of $N\simeq90$ atoms per lattice site at a temperature of T=2.5 $\mu$K. At this point we have reached PSD$\sim$0.1, and if we continue the cooling in this geometry, we observe that the PSD decreases due to the strong light-induced atom loss.

To prepare the atoms in a quasi-1D trapping geometry we proceed as follows: we adiabatically turn on a second lattice trap ($y$-trap) transverse to the first one with $w_y=6.5\ \mu$m waist and detuned from the $x$-trap by 160 MHz. This configuration creates a two dimensional array of elongated cigar-shaped traps along the $z$ direction (see Fig. \ref{setup}). This second lattice is created by a vertically polarized beam and its retroflection with polarization rotated by 70\degree. We use 5.5 mW of power to achieve trapping frequencies of $\omega_{y}=2 \pi \times 50$ kHz along the lattice and $\omega_{\perp y}=2 \pi \times 2.5$ kHz transverse to it, with a calculated trap depth of $U_y/h=0.47$ MHz. 

Before the $y$-trap turns on, the cold atoms in the $x$-trap are distributed at the bottom of the potential with a root-mean-square radius of 1.3 $\mu$m in the $y$-direction, so that most of the atoms are loaded into 3 lattice sites along the $y$-trap (see Fig. \ref{setup}a for reference to the coordinate system and  Fig. \ref{setup}c-d for the trap geometry). Immediately after switching on the $y$-trap, we adiabatically turn off the $x$-trap while increasing the power of the $y$-trap, allowing us to further compress the atoms in the $x$-direction, and to remove the atoms that are not confined to the overlap region of the two traps. The $y$-trap power is then increased by a factor of ten, producing a transverse frequency of $\omega_{\perp yc}=2 \pi \times 8$ kHz. At this point the temperature of the atoms has risen to about 20 $\mu$K, leading to a root-mean-square cloud size of 0.7$\mu$m along x. After 10 ms of thermalization the $x$-trap is adiabatically turned on. Finally, the $y$-trap power is adiabatically ramped back down to its previous value (Fig. \ref{setup}c). The entire process of compressing the atoms along both lattices takes 40 ms, without significant reduction of the PSD. At the end of this stage we have about 1000 atoms at 5 $\mu$K distributed in a 2D array with rms size $\sim1.3\times2.5$ lattice sites in the $x$ and $y$ directions, respectively, obtaining a peak occupation of $N\simeq50$ atoms per cigar-shaped trap.

The final cooling stage follows the same dRSC scheme as the pre-cooling stage, but now in two dimensions ($x$ and $y$). The trapping frequencies are $\omega_{x,y}=2 \pi \times 50$ kHz in the transverse directions and $\omega_{z}=\sqrt{\omega^2_{\perp x}+\omega^2_{\perp y}}=2 \pi \times 2.9$ kHz along the weakly confined vertical direction. After 200 ms of cooling, we reach a kinetic energy of the free expansion that corresponds to the ground-state kinetic energy of the tightly confined direction, i.e. we cool to the 2D ground state in the $xy$-plane. During cooling, atoms are lost at a moderate rate due to light-assisted inelastic collisions; once the atoms are cooled to the 2D ground state, the loss rate substantially reduces, presumably due to the lower cooling and associated optical pumping rates.

The loss during the cooling is due to two-body collisions, with more loss occurring in the traps in the central region containing initially more atoms. We simulate the atom number distribution in each trap at fixed total loss for the ensemble during the final cooling ($N_{\text{initial}}=1000$ atoms and $N_{\text{final}}=300$ atoms), and find that this leads to a rather flat distribution in atom number, with most traps containing $N_1=6$ atoms.

\begin{table}[]
\begin{tabular}{|l|l|}
\hline
Trap wavelength    & 1064 nm            \\ \hline
$x$-beam power  & 100 mW        \\ \hline
$y$-beam power  & 5.5 mW        \\ \hline
$x$-beam waist & 17 $\mu$m     \\ \hline
$y$-beam waist & 6.5 $\mu$m     \\ \hline
$\omega_{x,y}$   & $2\pi \times 50$ kHz \\ \hline
$\omega_{z}$       & $2\pi \times 2.9$ kHz \\ \hline
Trap depth $U$     & $h \times 2.7$~MHz            \\ \hline
Magnetic field $B$    & 0.15 G             \\ \hline
\end{tabular}
\caption{Experimental parameters.}
\end{table}

\subsection{Atom Number Distribution}

During the last stage of cooling the atoms are lost mainly due to light-induced collisions, which for a given laser intensity and detuning depend on the probability of finding two atoms near each other while one of the atoms is not in the dark state $\ket{F=3,m=3}$. We can model light-induced losses as a two-body process, where the reduction of the number of atoms is given by the solution of the differential equation $\dot{N}=-\alpha N^2$, namely
\begin{equation}
N(t)=\frac{N_0}{\alpha N_0 t+1}.
\label{eq:sm:atomnumber}
\end{equation}
Here $N_0$ is the initial number of atoms and $\alpha$ is light-assisted two-body loss rate. We model the density profile of the atoms remaining in the trap by starting with a Gaussian distribution and letting it evolve following Eq. (\ref{eq:sm:atomnumber}). Fig. \ref{SMfig1} shows the evolution during cooling of $N_{\text{initial}}=1000$ atoms in a Gaussian distribution with a width given by the geometric average of the rms size of the sample after compression in the x and y direction ($\sqrt{0.7 \mu\text{m}\times1.3\mu\text{m}}$), which corresponds to 1.8 lattice sites. By the time the atoms are cooled down to the transverse ground state there are about $N_{\text{final}}=300$ atoms. In this case we predict an almost flat density distribution of about $50$ tubes in two dimensions with $6$ atoms per tube.

\begin{figure}[h]
\includegraphics[width=0.6\textwidth]{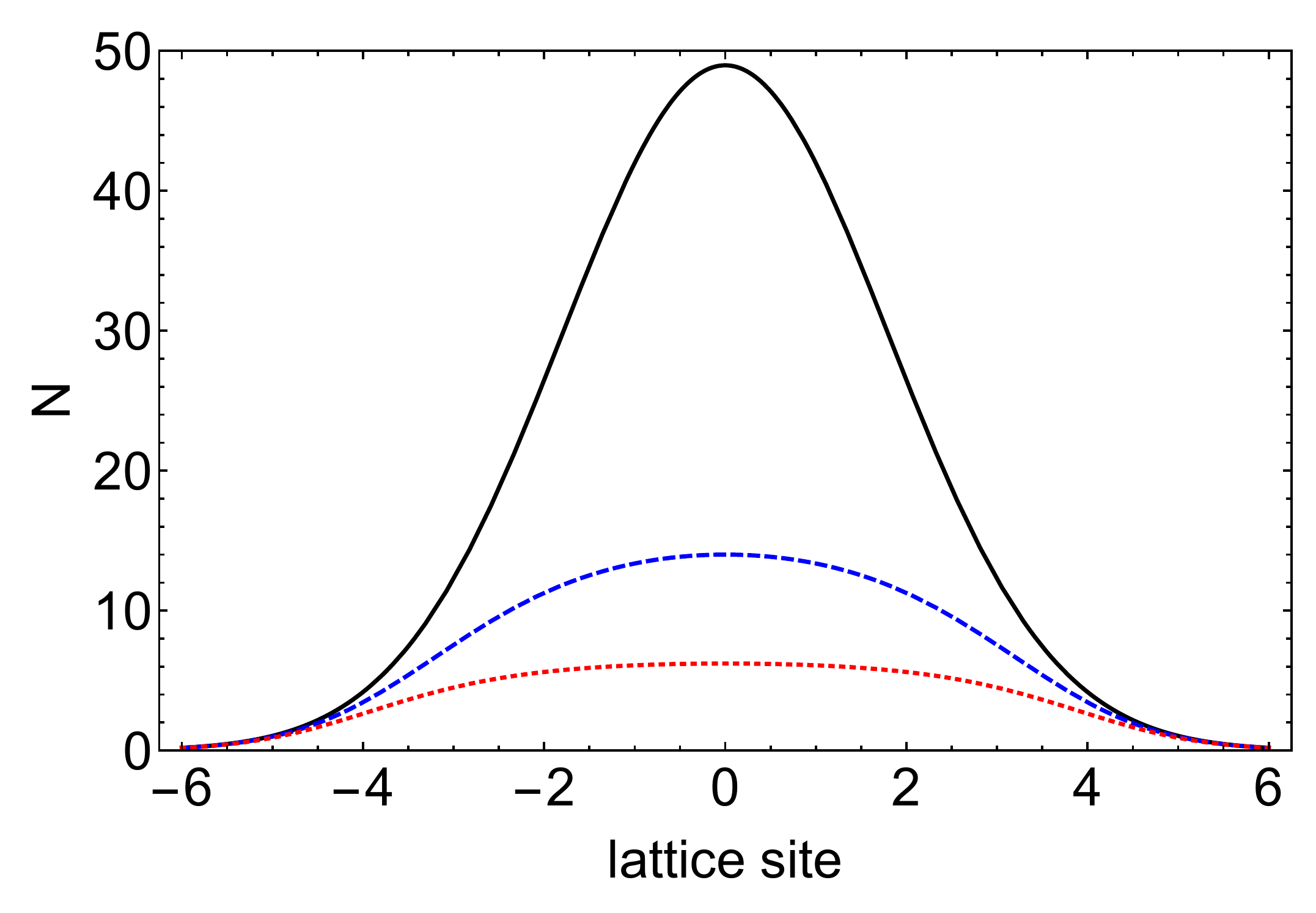}
\caption{Atom number distribution per lattice site at different cooling times. After loading into the 2D array of traps, the ensemble contains a 1000 atoms in a Gaussian distribution of the atom number per lattice site (solid black). As the atom number in each trap is reduced by two-body loss to a distribution with 500 atoms (dashed blue) and 300 atoms (dotted red), the nonlinearity of the loss leads to a flat distribution of atom numbers.}
\label{SMfig1}
\end{figure}

\subsection{Magnetic Field Dependence}

We apply a magnetic field rotated from the $y$ axis by a small angle $\alpha$. To optimize the optical pumping into the $|6\text{S}_{1/2},\text{F}=3,\text{m}_{\text{F}}=3\rangle$ state, we scan the angle $\alpha$ by minimizing the atom loss at large optical pumping power. For the thusly obtained angle $\alpha \approx 10^{\circ}$, Fig. \ref{SMfig2} shows the performance of dRSC versus magnetic field. We observe that the minimum temperature is reached near $B=150$~ mG, consistent with a Zeeman frequency splitting that equals the trapping frequency $\omega_{x,y}$ in the directions of tight confinement.

\begin{figure}[h]
\includegraphics[width=0.5\textwidth]{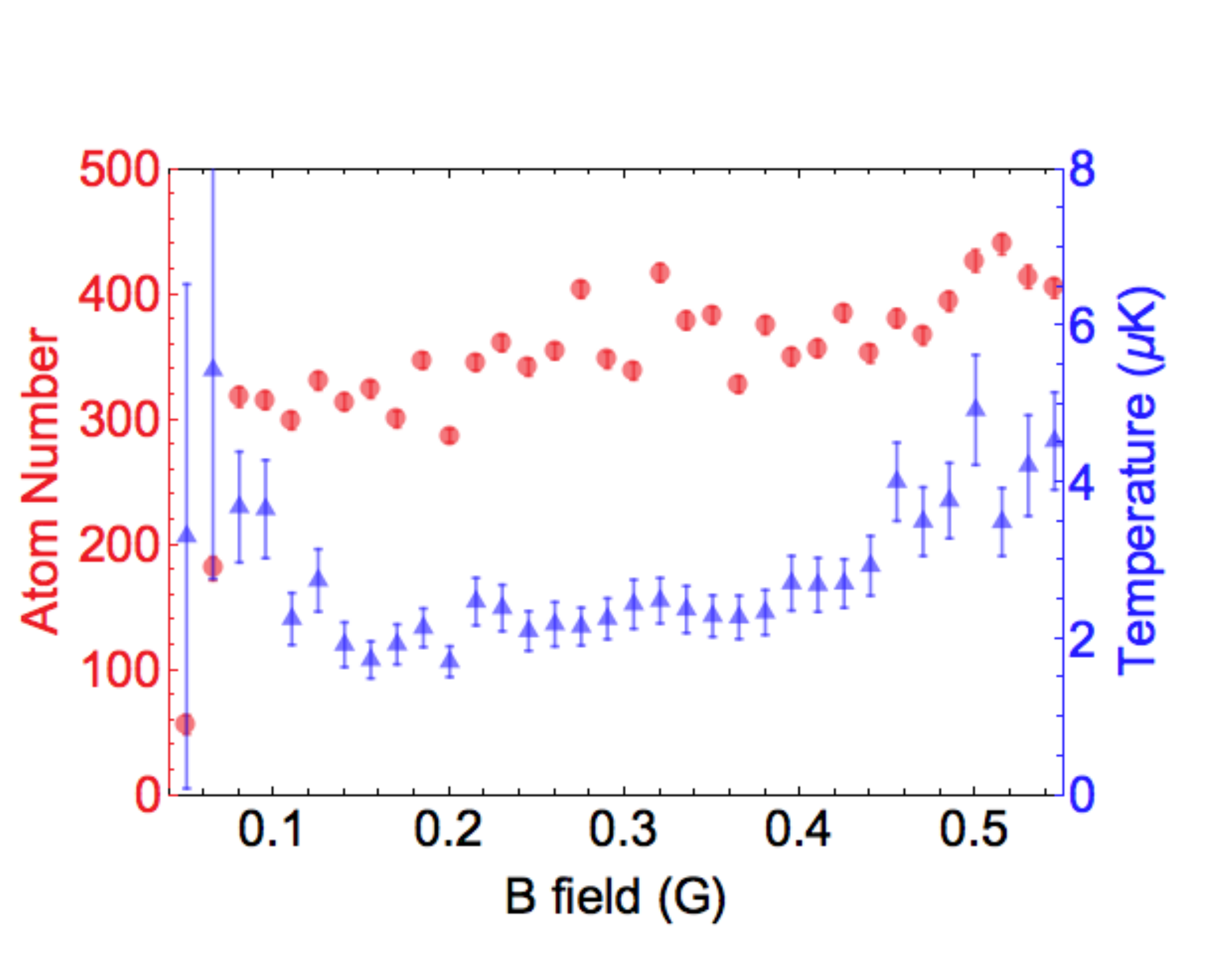}
\caption{Number of atoms (red circles) and temperature (blue triangles) as a function of magnetic field after 100 ms of cooling.}
\label{SMfig2}
\end{figure}

\subsection{Calculation of Phase Space Density}

The (classical) phase space density is defined as the probability for a single atom to populate the three-dimensional quantum ground state of the system, multiplied by the number of atoms per trap $N_1$,
\begin{equation}
\text{PSD}=N_1 P_0=N_1 p_{0,x}p_{0,y}p_{0,z},
\end{equation}
where $ p_{0,i}$ is the ground-state occupation along the $i$-direction, given by $p_{0,i}=1-e^{-\frac{\hbar\omega_i}{k_{B} T_i}}$. The kinetic energy observed in time-of-flight is half of the total energy and given by
\begin{equation} 
K_i=\frac{1}{2}\hbar\omega_i\left(\frac{1}{2}+\frac{1}{e^{\frac{\hbar\omega_i}{k_{B} T_i}}-1}\right).
\end{equation}
This leads to a relative ground state occupation of 
\begin{equation}
p_{0,i}=\frac{2}{\frac{4K_i}{\hbar\omega_i}+1},
\end{equation}
and a phase space density given by
\begin{equation}
\text{PSD}=N_1\prod_{i=x,y,z}\frac{2}{\frac{4K_i}{\hbar\omega_i}+1}
\end{equation}

\subsection{Characterization of Time-of-Flight Distributions}

We characterize the velocity distribution of the sample using time-of-flight measurements. We let the gas expand for 800 $\mu$s before an absorption image is taken. We integrate over the vertical or horizontal direction of the image to obtain the velocity distribution along the direction of tight or weak confinement, respectively. We fit a Gaussian distribution to all the data points and then eliminate all the points within one standard deviation of that fit (see Fig. \ref{SMfig3}). We then fit a Gaussian distribution to the remaining tails. In the direction of tight confinement we always observe a distribution that is well approximated by a Gaussian. On the other hand, the momentum distribution along the direction of weak confinement has a non-Gaussian central part. We consider these characteristic non-Gaussian momentum distributions as a signature of a quantum degenerate gas. We fit the non-Gaussian fraction of the data to a Thomas-Fermi distribution (inverted parabola) with reasonable agreement, although the exact momentum distribution near zero-momentum in general does not have an analytic functional form. However, we can quantify the fraction of atoms that do not follow a thermal distribution by the ratio of the area under the inverted parabola to the total area.

\begin{figure}[h]
\includegraphics[width=0.8\textwidth]{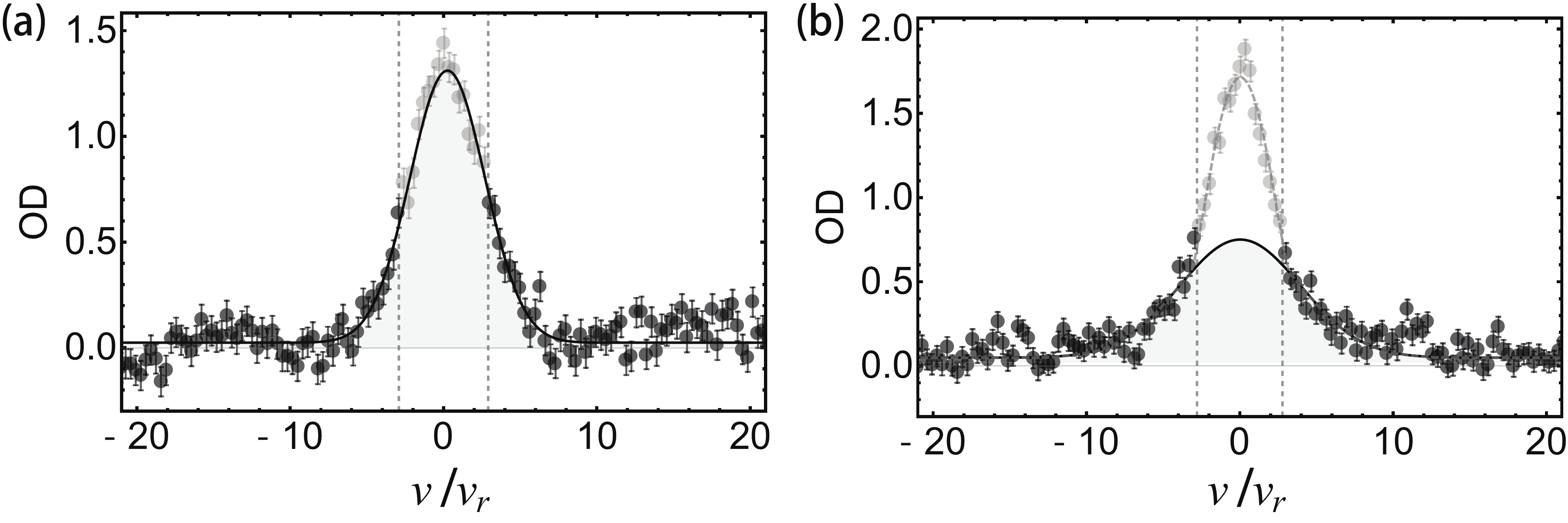}
\caption{Velocity distribution of the atoms normalized by the recoil velocity. (a) Velocity distribution along the tightly confined direction, fit to a Gaussian. (b) Velocity distribution along the weakly confined direction. The vertical dashed lines correspond to plus and minus one standard deviation. The solid black curve is the fit to a Gaussian distribution considering only the darker data points and the gray dashed curve in (b) is a Gaussian distribution plus an inverted parabola.}
\label{SMfig3}
\end{figure}

\subsection{Photo-induced Losses and $\mathbf{g^{(2)}}$ Correlation Function}

In the regime where the loss and temperature change are small, i.e., the average density $\aver{n}$ is approximately constant, we can describe the atom number loss due to photo-induced two-body inelastic collisions as an exponential decay determined by the differential equation
\begin{equation}
    \dot{N}=-\Gamma N.
    \label{eq:SM:2b}
\end{equation}
The loss rate $\Gamma=G g^{(2)}\langle n\rangle$ depends on the three-dimensional average atomic density $\aver{n}$, the photo-association rate constant $G$, and normalized probability to find two atoms at vanishing interatomic distance $g^{(2)}(\mathbf{r}=0)$.

The atom-atom correlation function $g^{(2)}$ can be evaluated by comparing densities and decay rates for the cases of one- and two-dimensional gases, by turning on or off one of the trapping lattices. The ratio of $g^{(2)}$ in both cases is
\begin{equation}
  \frac{ g^{(2)}_{\text{1D}}}{g^{(2)}_{\text{2D}}}=\frac{ \Gamma_{\text{1D}}}{\Gamma_{\text{2D}}}\frac{ \langle n\rangle_{\text{2D}}}{\langle n\rangle_{\text{1D}}}\frac{ G_{\text{2D}}}{G_{\text{1D}}},
\end{equation}
where the densities can be calculated from the measured atomic temperature and the trapping frequencies. We fit the data to the solution of Eq. (\ref{eq:SM:2b}) to extract the value of $\Gamma$. The quantity $ \Gamma_{\text{1D}}\langle n\rangle_{\text{2D}}\Big/\Gamma_{\text{2D}}\langle n\rangle_{\text{1D}}$ is shown in Fig. \ref{lossrate}a.  If we assume that the two-dimensional gas is approximately thermal, for which  $g^{(2)}_{\text{2D}}=2$, we obtain 
\begin{equation}
g^{(2)}_{\text{1D}}\simeq0.1\frac{G_{\text{2D}}}{G_{\text{1D}}}.
\end{equation}
In an ideal experiment, the photo-association rates are constant, allowing to extract the value of $g^{(2)}$ for different trap geometries, but in practice they may differ. One possibility explanation for having $G_{\text{1D}}\neq G_{\text{2D}}$ are the different heating and cooling rates in both configurations, however we measure both rates to be similar in 1D and 2D. We hypothesize that the main mechanism for having $G_{\text{1D}}\neq G_{\text{2D}}$ is the reduction of on-resonance photo-association due to the tight confinement. Two atoms photo-associate with resonant light if they are separated by a distance close to the Frank-Condon point \cite{Burnett1996}. In the case of a 1D trap this holds true only for atoms along the z direction, since the transverse confinement is smaller than the Frank-Condon point. In 2D, the resonant condition is satisfy for atoms in a circle in the two dimensional plane. The ratio $G_{\text{2D}}/G_{\text{1D}}$ is of the order of the ratio of the Frank-Condon point $r_{\text{C}}\sim \bar{\lambda}=135$~nm and the transverse confinement of the trap $a_{\perp}=39$~nm. Considering $G_{\text{2D}} \sim (r_C/a_{\perp})G_{\text{1D}}$, we get
\begin{equation}
g^{(2)}_{\text{1D}}\sim 0.4.
\end{equation}
For our parameters, the temperature $T$, normalized to the degeneracy temperature $T_\text{D}=\hbar^2 n_{1D}^2/(2 m k_B)=46$~nK of the repulsive Lieb-Liniger model \cite{Kormos2011}, is $\tau=T/T_{\text{D}}=26$. For this temperature and $\gamma=8$ we expecct $g^{(2)}_{\text{1D}} \approx 0.4$ \cite{Kormos2011}.

\subsection{Three-Body Losses and $\mathbf{g^{(3)}}$ Correlation Function}

The density evolution of an atomic ensemble in a trap with three-body loss is governed by
\begin{equation}
 \dot{n}=-K n^3,
 \end{equation}
 where $K$ is the three-body loss rate coefficient, and $n$ is the local density. This equation can be re-written in terms of total atom number as
 \begin{equation}
 \dot{N}=-C N^3,
 \end{equation}
where for a harmonic trap the coefficients are related by
\begin{equation}
K=\frac{3^{3/2}}{\rho_{\text{3D}}^2}C,
\end{equation}
where $\rho_{\text{3D}}$ is the single-atom peak density in a trap given by 
\begin{equation}
\rho_{\text{3D}}=\frac{1}{(2\pi)^{3/2}x_0y_0z_0},
\end{equation}
where $x_0$, $y_0$, and $z_0$ are the root-mean-square size of the atomic distribution in each individual trap given by the temperature and the trapping frequencies.

The solution of the differential equations is
\begin{eqnarray}
N(t)=\frac{N_{1}}{\sqrt{2 C N_{1}^2 t+1}},
\end{eqnarray}
with $N_{1}$ being the initial number of atoms in the trap. Since all traps are almost equally filled due to the two-body loss during the preparation, the evolution of total atom number is governed by the same equation. This is the expression we use for our fits to obtain the three-body loss rate (Fig.  \ref{lossrate}b).

The three-body loss rate is proportional to the overlap of the wavefunction of three atoms, characterized by the correlation function $g^{(3)}(\mathbf{r}=0)$. We consider $K=K_0g^{(3)}$ and compare the cases for one- and two-dimensional gases, assuming that $K_0$ is a constant, independent of the dimensionality of the problem. If we also assume that the two-dimensional gas is approximately in a thermal distribution, where $g^{(3)}_{\text{2D}}=6$, then our measurements of three-body loss yields 
\begin{equation}
    g^{(3)}_{\text{1D}}\simeq0.05.
\end{equation}

Analogously to $g^{(2)}$, $g^{(3)}$ can be calculated from theory even at finite temperature, and for our parameters ($\gamma=8,\tau=22$) is given by $g^{(3)}_{\text{1D}}\simeq 0.06$ \cite{Kormos2011}.

The main source of systematic errors for measuring $g^{(3)}_{\text{1D}}$ comes from the uncertainty of the atomic density in the case of a 2D traps (1D lattice). We observe neighboring traps getting populated while atoms are held in the dark (without cooling). This effects decreases the density of atoms by a factor of 4, and adds a uncertainty of 50 percent in determining the value of $K_{\text{2D}}$.

\end{widetext}
\end{document}